\documentclass[5p,times,review]{elsarticle}
\usepackage{graphicx}
\usepackage{pifont}
\usepackage{natbib}
\usepackage{geometry}
\usepackage{fleqn}
\usepackage{txfonts}

\usepackage{booktabs}

\usepackage[caption=false,font=footnotesize]{subfig}

\usepackage{amsfonts}
\usepackage{amsmath}
\usepackage{amssymb,mathrsfs}
\usepackage{array}
\usepackage{flafter}

\usepackage{verbatim}

\usepackage{color}
\usepackage{psfrag}
\usepackage{umoline}
\usepackage{hyperref}
\usepackage{indentfirst}

\appendix

\allowdisplaybreaks[1]
\makeindex             

\usepackage[T1]{fontenc}
\usepackage[utf8]{inputenc}

\usepackage{lipsum}

\usepackage[figuresright]{rotating}

\begin{document}

\begin{frontmatter}
\title{A Predictive Model for Oil Market under Uncertainty: Data-Driven System Dynamics Approach }
\author[rvt]{Sina Aghaei\corref{cor1}}
\cortext[cor1]{Corresponding author}
\address[rvt]{Department of Industrial \& Systems Engineering, University of Southern California, Los Angeles, CA 90007, USA}
\ead{saghaei@usc.edu}
\author[focal]{Amirreza Safari Langroudi}
\address[focal]{School of Management, Sabanci University, Istanbul 34956 , Turkey}
\author[els]{Masoud Fekri}
\address[els]{School of Industrial Engineering, Iran University of Science and Technology, Tehran 16846-13114 , Iran}

\begin{abstract}
In recent years, there have been a lot of sharp changes in the oil price. These rapid changes cause the traditional models to fail in predicting the price behavior. The main reason for the failure of the traditional models is that they consider the actual value of parameters instead of their expectational ones. In this paper, we propose a system dynamics model that incorporates expectational variables in determining the oil price. In our model, the oil price is determined by the expected demand and supply vs. their actual values. Our core model is based upon regression analysis on several historic time series and adjusted by adding many casual loops in the oil market. The proposed model in simulated in different scenarios that have happened in the past and our results comply with the trends of the oil price in each of the scenarios. 
\end{abstract}
\end{frontmatter}
\section{Introduction}\label{sec:intro}

In recent years, there have been multiple efforts to model the oil market to capture its volatile behavior. In most of the models, both supply and demand side have different dependencies of the consumers on the oil price. However, in almost all of the traditional models, the actual value of the supply and demand is used for modeling the oil market \cite{naill1992system, kilian2012agnostic, roe2001us}. However, due to the failure of the traditional models in capturing price trend, \cite{rafieisakhaei2016analysis, lean2010market, hallock2004forecasting} started to use expectational values instead. The accuracy of the model proposed in \cite{barazandeh2016effect} shows the validity of the expectational models.
     
The emerging markets like Russia, India, Brazil, and China (RIBC) are heavily thirsty of oil and their industries are getting even more dependent on oil consumption \cite{13_financialtimes_2015, basher2006oil, morck2000information}. However, in the European Union and United States (EUS) there are usually policies on limiting the pollution and air emission which directly affect the amount of fuel consumption and the standards that are used in the manufacturers like car producers \cite{rafieisakhaei2016modeling}. Moreover, most of the researches on oil substitutes are performed and enforced in the EUS countries. Therefore the variables on different sides of the demand involve different types of dependencies on the oil price. It should be noted that the two sides of the demand chain (EUS and RIBC) are actually interrelated, as well \cite{jaffe2002beijing}.

Therefore, the factors enforcing the oil price are more than just the aggregated demand and supply. For instance, after the OPEC meeting on November 2014, the oil price gained a much higher speed in reduction \cite{baumeister2015understanding}. This was because, unlike the expectations, the OPEC producers decided to keep the production rate unchanged, where as the oil price had already fallen by \$20 per barrel since June 2014. Or as another instance, one could think of the new types of energy resources that have been explored. For instance, as a result of the Shale boom, the US is now the largest producer of the natural gas \cite{encykey} which puts even a greater pressure on the oil price \cite{economides2009state}.

In this paper, we extend the proposed model in \cite{rafieisakhaei2016supply} to capture some other parameters and variables that are specially involved in determining the oil price and the expectations that shape it other than just the demand and supply. First, we make a cause and effect model based on the casual relationships among the variables. Next, we add the formula and build the stock and flow SD model upon the causal one. Similar to \cite{kilian2009not, krichene2002world}, we separate the total oil demand to three different categories of the demand by EUS, RIBC and the other countries. Likewise, the supply side is categorized as OPEC, US, spare oil supply, smuggled oil production, and the production by other countries. Our model incorporates the new expectational parameters. Particularly, the two variables ``Expectation on Demand Side" and ``Expectation on Supply Side" summarize effects of the unconventional factors \cite{Raf15}. These variables play a vital role in our model. They are amplifier or shrinker of the total supply and demand in this model. We build the model, its causal loops, and the mathematical formula in section \ref{sec:Model, Loops and Formula}. Similar to \cite{azadeh2016unique}, we use real data of oil price, demand, supply and economic growths for building the mathematical relations  of the core model using the regression analysis on key variables. Then, in section \ref{sec:simulations}, we provide our simulation results for five scenarios and compare the results with the time-wise corresponding real data. We show that our results successfully comply with the real data's trends in each of the scenarios that have happened in the past. Moreover, we provide several cases that try to predict the oil price in a 60-day time window for some possible scenarios. Finally, section \ref{sec:conclusion} concludes the paper.

\section{The model structure}\label{sec:Model, Loops and Formula}
In this section, we describe details of the proposed model. In the first section, we will explain the supply side of the model and the second part describes the demand section. Details of the models is described in the following. The proposed model could also be modified to capture the volatility in car market \cite{barazandeh2016effect}.

\subsection{The Supply Loop}
\begin{figure}[ht!]
  \centering
  {\includegraphics[width=3.5in]{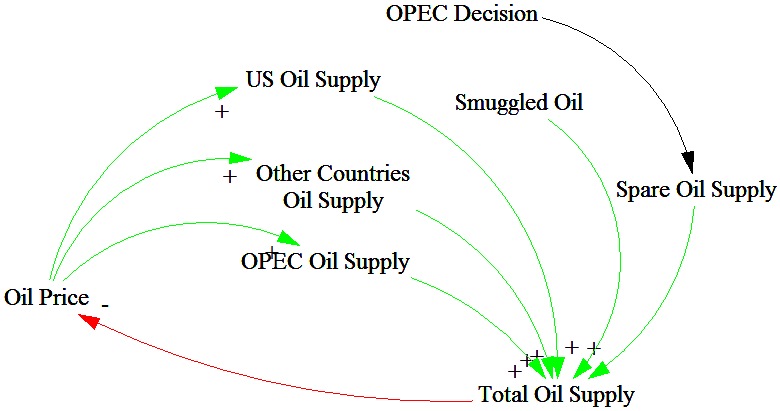}}
  \caption{The main supply loop and the sign of its arrows.\label{fig:Supply_Main_Loop} }
\end{figure}

In this subsection, we provide the main causal loops on the supply side of the chain. Figure \ref{fig:Supply_Main_Loop} depicts this main loop whose parameters and variables are explained next.

\textit{Main Components of the Supply:} The expected total oil supply consists of the `OPEC Oil Supply ($ S_{OPEC} $)', `US Oil Supply ($ S_{US} $)', `Other Countries Oil Supply ($ S_{Other} $)', `Smuggled Oil Supply ($ S_{Smuggled} $)', and `Spare Oil Supply ($ S_{Spare} $)'. The `Smuggled Oil Supply' models the production of the oil in the regions with political upsets that the governments of the producing countries do not have the political controls over the country \cite{7_africaecon.org_2015, gately1984ten}. For instance, in 2014, oil production in Syrian and Iraqi regions that were out of the control of their governments was approximately 600,000 barrels a day \cite{11_almonitor_2015,herbert2014partisans}. After political troubles in that region, the expectations favored a reduction in the oil production in the corresponding countries for long-run. However, this only happened for a short period of time just before the market was overwhelmed by the smuggled oil production which accounts for less than one mb/d. However, this is a considerable amount of production in the oil market, where only one mb/d surplus in the production can cause the prices to plunge sharply. We express the mathematical relations between $ S_{Smuggled} $, $ S_{Spare} $ and the parameter `OPEC's Decision', which models the effect of OPEC's decisions in the oil market, in the next subsection where we talk about Geopolitical Upsets and the Expectation on Supply. Moreover, we have separated the producing countries' shares in the oil market. This is mainly due to the fact that the production resources in different countries differ from each other. Therefore, the Expected Total Oil Supply can be written as follows:
\begin{align}
TOS = S_{OPEC}+S_{US}+S_{Other}+S_{Smuggled}+S_{Spare}.
\end{align}

\subsection{The Expectation on Oil Supply Loops}
\begin{figure}[ht!]
  \centering
  {\includegraphics[width=3.5in]{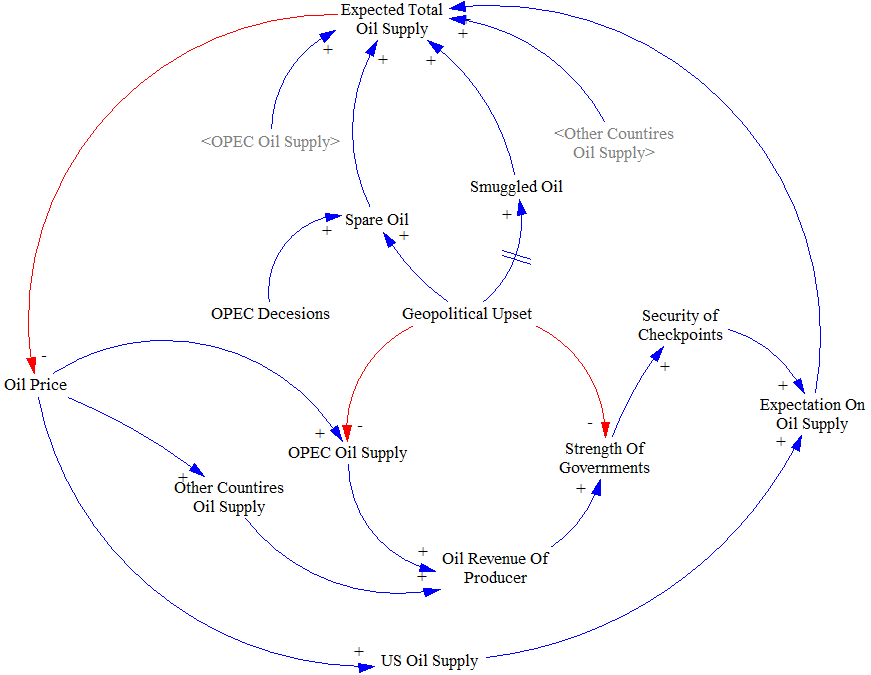}}
  \caption{The main expectation on supply loop and the sign of its arrows.\label{fig:Expectation_on_Supply_Loop} }
\end{figure}

In this subsection we provide factors that constitute the `Expectation on the Oil Supply ($ ExS $)'. Figure \ref{fig:Expectation_on_Supply_Loop} depicts the main loops and causal relations in determining the value of this variable.

\textit{The Expectation on the Oil Supply:} The $ ExS $ variable models the factors that form the expectations in the changes of the oil supply. Particularly, in the occurrence of some event that affects the expectation on the direction of future changes of supply, these expectations eventually are transformed through some variables and parameters into actual changes in the oil price. In such cases, although the actual change in the oil supply might not have occurred yet, the oil price changes accordingly as if the supply has actually changed in the predicted direction. This is why many of the variables that are involved specifically in the expectation loops are modeled such that they get influenced through the direction of changes of the main variables. For instance, some of them change according to the smoothed derivatives of main variables such as oil revenues or economic growth. Because indeed, in practice, the expectations are shaped according to the (recent) direction of changes of the definitive economic factors. Thus, in a novel way, we have modeled these kinds of parameters to affect the oil price only through the oil supply. The $ ExS $ is a dynamic coefficient that acts as an amplifier or suppressor on the global oil supply. Thus, the oil price is not changed directly, and rather, the expectation on supply changes the supply value artificially, and it is only through the supply to demand ratio that the oil price changes. In other words, it is as if the oil supply has really changed and the oil price takes appropriate response to such a change. Finally, the actual supply is still shown in the model through the variable `Actual Total Supply', which is the `Expected Total Oil Supply ($ TOS $)' divided by the $ ExS $. 

\subsection{The Demand Loop}
\begin{figure}[ht!]
  \centering
  {\includegraphics[width=3.5in]{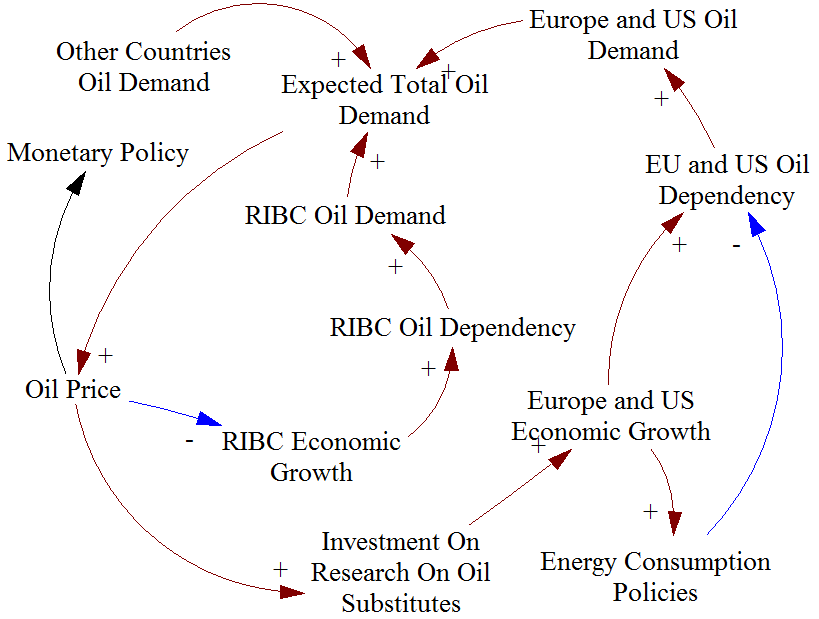}}
  \caption{The main demand loop and the sign of its arrows.\label{fig:Demand_Main_Loop} }
\end{figure}

In this subsection, we provide the causal loops that determine the total amount of global oil demand. Figure \ref{fig:Demand_Main_Loop} reflects the main loops that are effective in the demand side of the chain.

\textit{Main variables on Demand:} The `Expected Total Oil Demand ($ TOD $)' is determined by the `EU and US Oil Demand ($ D_{EUS} $)', `Russia, India, Brazil, China Oil Demand ($ D_{RIBC} $) and the `Other Countries Oil Demand ($ D_{Other} $)'. The reason to separate RIBC oil demand is that these countries have a growing economy that is heavily thirsty of oil \cite{birol2010world}. Therefore, their industries are increasingly consuming oil and this makes it slightly different from the EU and US side, where the oil demand growth rate is much lower. Particularly, through new polices on oil consumption or even finding substitute energy resources, they even tend to decrease their oil dependencies in the future. Therefore, structure of the oil demand and even the economic growth is different on the two sides of demand. Thus, there is a need to model their causal loops in slightly different ways with different variables and functions. Thus we can compute $ TOS $ as:
\begin{align}
TOD = D_{EUS} + D_{RIBC} + D_{Other}.
\end{align}
Throughout this paper, by total oil demand we refer to Expected Total Oil Demand, unless explicitly stated.

\textit{Energy Resources and Policies on Oil Consumption:} The variable `Policies on Oil Consumption' reflects the effects of policies and regulations on the oil-dependent industries like car factories. Generally, due to the factors such as air pollution caused by burning the fossil fuels, limitations on the oil reserves \cite{sterman1988modeling}, and vulnerability of the economic growth caused by the oil price, there is a tendency to reduce economy's dependency on the oil consumption \cite{18_union_of_concernedscientists_2015}. For instance, the newer cars have higher MPGs (Miles Per Gallon). Therefore, we have modeled the effect of those regulations, which are themselves a functions of the economic health, with the aforementioned variable. Moreover, the variable `Other Energy Resources' reflects the effect of oil substitutes such as natural gas that have been growingly used by the EU and US countries. Particularly, since the Shale Boom, the US has been the first producer of the natural gas in the world \cite{19_mccain_2015,encykey}. This has also put higher pressures on the oil price which was very volatile in the second half of 2014, and seems to be onwards (in 2015) \cite{1_barton_2015,2_webb_2015}. As another instance, the lithium battery powered cars are being considered as the new trend in the car industry which also reduces the oil dependency \cite{20_conger_2015}, if the electrical energy that is used to charge those batteries are themselves produced by the renewable energy resources or other types of fossil fuels. These are some of the variables that are specially important on reducing the oil dependency of the EU and US countries. Unfortunately these kinds of policies are sometimes overlooked or less considered in RIBC countries. This is the reason that we have not included those in that side of the model.

\subsection{The Expectation on Demand Loops}
\begin{figure}[ht!]
  \centering
  {\includegraphics[width=3.5in]{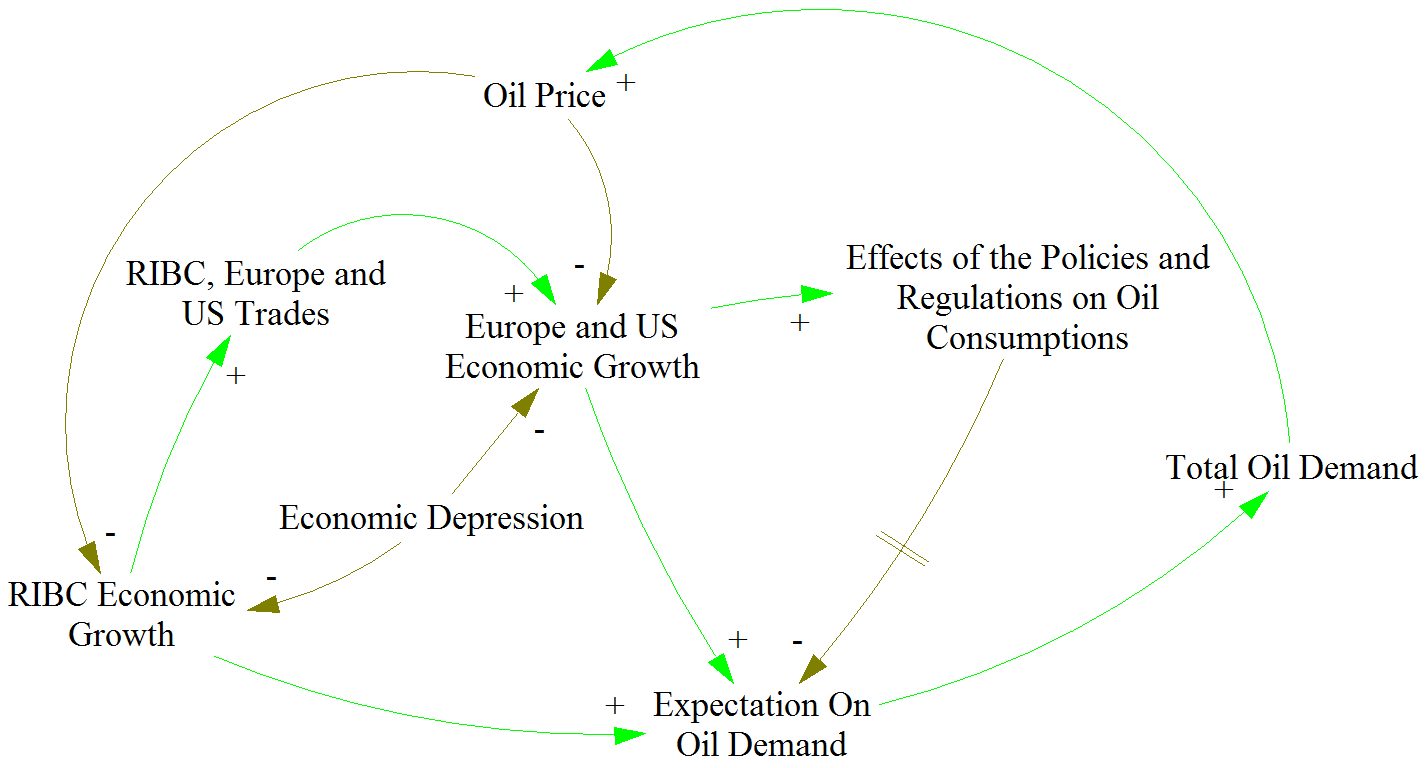}}
  \caption{The main expectation on demand loop and the sign of its arrows.\label{fig:Expectation_on_Demand_Loop} }
\end{figure}

In this subsection, we provide the loops that are important in forming the Expectation on Oil Demand. The main loops are depicted in Fig. \ref{fig:Expectation_on_Demand_Loop}. The rest of this section elaborates the structure of the dependencies in these loops.

\textit{The Expectation on Oil Demand:} Likewise the supply chain, the $ ExD $ variable models the factors that form the expectations in the changes of demand. To emphasize the need for this variable, consider a circumstance where the International Monetary Fund (IMF) declares the expected economic growth of EU, US or China \cite{imf2015world}. Usually, whenever these kinds of declarations happen, the oil price reacts to them. This is because the expected economic growth, as discussed in the previous subsection, is interrelated with the global oil demand. Therefore, an expectation on the economic growth rate is translated through some variables and parameters into an expectation over the global oil demand. In such a case, although the actual change in the oil demand might not have occurred yet, the oil price changes slightly as if the demand has actually changed in the predicted direction. Therefore, one of the contributions of this paper is to model those kind of parameters in determining the oil price. Likewise supply, $ ExD $ is a dynamic coefficient that acts as an amplifier or suppressor on the global oil demand. Therefore, once again, instead of directly changing the price, the expectation on demand, changes the demand value artificially and it is only through the demand that the oil price changes. In other words, it is as if the oil demand has slightly changed and the oil price responds to that correspondingly. Notice that, the `Actual Total Demand' is the `Expected Total Oil Demand' divided by $ ExD $.

\textit{Economic Depression:} One of the main ongoing economic events since 2008 has been the economic depression. Although US economy is said to have recovered from the depression in 2015 \cite{15_bartash_2015}, some of the economies in the European Union are still suffering from the effects of this crisis. This is one of the factors that has a direct impact on the economic growth, for instance, through unemployment rate, exports, gross domestic product, etc. Moreover, through the economic growth change rate data, it can be understood that this factor is indeed important in forming the expectations on the demand change rate, whence the oil price \cite{14_marketrealistcom_2015}. In our model, the `Economic Depression ($ EconDep $)' is an initial coefficient that acts on the loops containing the economic growth and the expectations on demand to reflect the adverse effects of the depression on the global oil demand. Particularly, in the initiation of Economic Depression, $ ExD $ decreases, which causes the price to reduce. This can help recovering the growth, however, it is only until a long-term recovery of the economy that the demand increases back. Therefore, we have used a $ PULSE(start, duration) $ to model the $ EconDep $ and to trigger the loops that involve $ EconDep $ for the duration of time that we want to model the existence of depression. 

\textit{Effects of the Policies and Regulations on Oil Consumptions:} Although the policies and regulations have been considered in determining the actual amount of the oil demand with a great time constant, it is necessary to consider their effects on forming the expectations on demand with a lower time constant, as well. Particularly, on one hand, the regulations might take a much greater period of time to actually change the amount of demand by changing the oil-dependency \cite{22_the_economist_2015}. On the other hand, just as the polices get passed in the parliaments or organizations, they form the expectations on the demand change rate. However, their effects through this loop might be lower than the aforementioned side.

Now that we have elaborated the causal loops in the model, we are ready to simulate the model and bring the results corresponding to several different scenarios. The next section talks about these scenarios in detail.

\section{Simulations and Scenarios}\label{sec:simulations}
\begin{figure}[!t]
\centering
 \includegraphics[width=3.5in,height=6cm]{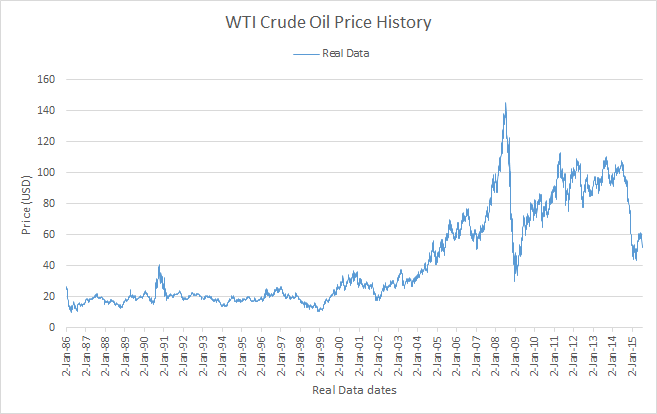}
  
\caption{WTI oil price history.}
  \label{subfig:WTI_history}
\end{figure}

\begin{figure}[!t]
\centering
 \includegraphics[width=3.5in,height=6cm]{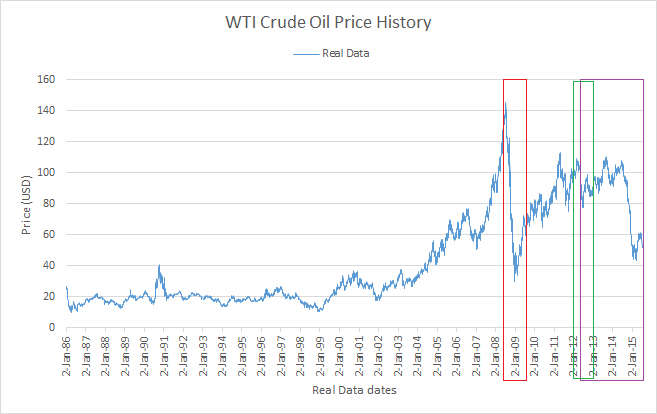}
  
\caption{Parts of the WTI oil price history that we have used in our simulations, shown with rectangles.}
  \label{subfig:WTI_history_marked}
\end{figure}

In this section, we evaluate the accuracy of the proposed model in different scenarios. Our simulations involve the cases where the parameters on the expectation on either demand or the supply side changes and the results are reflected in the oil price. In our simulations we have considered scenarios that are consistent with the real world scenarios and have used the West Texas Intermediate (WTI) index as our resource for the oil price real data. The WTI crude oil price history is shown in Fig. \ref{subfig:WTI_history} with the data taken from \cite{1_company_2013}. Moreover, in Fig. \ref{subfig:WTI_history_marked} we have shown the parts of the data that we have used in our simulations with three rectangles. Moreover, as will be explained later, in scenario A, we have used the data from 30th of Dec. 2011 to 10th of Jan. 2012, that we have not shown in this figure for the sake of cleanness of the figure. Moreover, in scenario E, we do a simulation for the period of time starting from Jan. 2012 until July 2015. Finally, in the last scenario, we provide future predictions of the price changes in several possible cases.


\subsection{Scenario A: Sanctions on the Production}
In this subsection, we simulate a scenario where the sanctions on oil producers have caused a 
reduction in their production.

\textit{International Sanctions:} The international society can decide to establish economic, political or other types of sanctions on the oil producing countries due to various reasons. There are many instances of those cases over the past 50 years. For instance, as mentioned before in the 90s, there were heavy economical sanctions on Iraq which had infected their economy with lots of problems. Over the past three years (since 2012), there has been many economical sanctions on Iran (as an OPEC member), which have limited their oil sales and production. We have used the real data of oil price for the year 2012, where these sanctions were one of the main reasons of the oil market's volatility. These kinds of sanctions have many adverse effects on the economies of the sanctioned countries. We have modeled these effects to gradually reduce oil production of the sanctioned countries using the real data, and have reflected the results of introduction of sanctions on one of the OPEC producers in Fig. \ref{subfig:Scenario_C_vensim}. In the simulation, we have assumed a placement of sanctions on one of the producers (at time 1) whose effects leads to a shortage in the supply with a reasonable delay (due to the stock and flow nature of the model), consequently giving an increasing rate to the oil price. Moreover, the OPEC Decision is still zero which simulates a case that the OPEC countries have not yet decided on compensating the supply shortage, which is consistent with the real data in 2012. Rationally, it is expected that these sanctions should cause the price to face a gradual increase. This is mainly due to the fact that the sudden reduction in oil production of one the countries might not be compensated by the usual growth of the oil supply. However, this is not what happens in reality. As it is seen in Fig. \ref{subfig:Scenario_C_vensim}, the oil price keeps its trend for slightly decreasing. This is mainly due to the fact that the US oil production has increased with a noticeable rate during the same period in 2012. Therefore, although, the OPEC oil production has reduced due to the reduction of oil production in Iran, the US oil production has compensated the same amount and even, by the end of 2012, the total oil production in the world has increased. This is indeed reflected in the figure and our simulations predict the oil price pretty well in this scenario, as well. The only slight increase occurs in the first immediate months that follow the imposition of the sanctions. However, this volatility is compensated and eliminated in the following months, and this is reflected in the figure, as well. The mechanism for these reductions and growths is as explained in the previous section through the supply and its corresponding expectation loops.

\subsection{Scenario B: Economic Depression}
In this subsection, we simulate effects of an economic depression in oil consuming countries.

\textit{Lower oil demand due to depression:} Figure \ref{subfig:Scenario_D_vensim} reflects the fluctuations of the oil price in a case where an economic depression has happened, specially in the EU and US. As shown in this figure, the oil price decreases in this case, due to a reduction in the demand. However, we assume that the depression starts to get better after 210 days, which causes the demand and subsequently the oil price to increase. We have used the year-long real data for the period of time from 30th of May, 2008 to 30th of May 2009. In the first months after the economic depression, the demand for oil reduces sharply in the EU and US side, which causes the oil price to fall down, however, it is the increased volatility that makes the prices dive from about \$133 in 30th of May, 2008 to about \$40 by the end of 2008. As mentioned before, this volatility is reflected in our model through the expectation variables, and indeed it is the effect of the better expectations in the beginning of 2009 (in addition to the increased demand of the RIBC), that contributes in increasing the oil price in 2009. Note that in the beginning of 2008, the oil price had seen sharp increases which had various cause and reasons. Some of these reasons were the strikes and unexpected events in the oil producing countries like Nigeria, Scotland and Canada \cite{hamilton2009causes,1_smith_2015}. Therefore, the growth of oil price was mainly due to the volatility in the supply and expectation on the future of the supply side of the market. In the later months of the 2008, some of these volatility sources were still existent which can be seen from the supply data in those days. Therefore, we have modeled those supply volatilities in the supply loops and the overall result is shown in Fig. \ref{subfig:Scenario_D_vensim} which shows that our simulation results can follow the oil price trend in the reflected period of time.

\section{Conclusion and Future Work}\label{sec:conclusion}
In this paper we proposed a model for oil market. The novelty of the model is in modeling the supply and the demand loops in which instead of the actual values, we used the expected values of supply and demand in determining the ratio. Our approach was to build a Stock and Flow model based on a cause and effect model containing the causal relations among various determinants, and train the raw model with the historic real data of price, demand and supply to make a core model that explains the price changes in normal situations. Then, based on various events that included unexpected oil price changes, we adjusted the model and its related loops that are active in each of those scenarios. Our model was successful in predicting the past trends of the oil price in both short and long term examples on the past eight years. The future models may include more expectational variables . Moreover, there are some specific reasons for the oil price changes that has influenced the market in the past. For instance, extensive use of computerized marketing in 2011 caused a decay in the oil price towards the middle of that year . Similarly, factors like US dollar strength, stock market indices (that get affected by many other commodities), and big political decisions are some of the other parameters that might be considered more extensively in future.
\bibliographystyle{elsarticle-num}

\bibliography{V2}

\begin{thebibliography}{10}
\expandafter\ifx\csname url\endcsname\relax
  \def\url#1{\texttt{#1}}\fi
\expandafter\ifx\csname urlprefix\endcsname\relax\def\urlprefix{URL }\fi
\expandafter\ifx\csname href\endcsname\relax
  \def\href#1#2{#2} \def\path#1{#1}\fi

\bibitem{naill1992system}
R.~F. Naill, A system dynamics model for national energy policy planning,
  System Dynamics Review 8~(1) (1992) 1--19.

\bibitem{kilian2012agnostic}
L.~Kilian, D.~P. Murphy, Why agnostic sign restrictions are not enough:
  understanding the dynamics of oil market var models, Journal of the European
  Economic Association 10~(5) (2012) 1166--1188.

\bibitem{roe2001us}
B.~Roe, M.~F. Teisl, A.~Levy, M.~Russell, Us consumers’ willingness to pay
  for green electricity, Energy policy 29~(11) (2001) 917--925.

\bibitem{rafieisakhaei2016analysis}
M.~Rafieisakhaei, B.~Barazandeh, M.~Tarrahi, et~al., Analysis of supply and
  demand dynamics to predict oil market trends: A case study of 2015 price
  data, in: SPE/IAEE Hydrocarbon Economics and Evaluation Symposium, Society of
  Petroleum Engineers, 2016.

\bibitem{lean2010market}
H.~H. Lean, M.~McAleer, W.-K. Wong, Market efficiency of oil spot and futures:
  A mean-variance and stochastic dominance approach, Energy Economics 32~(5)
  (2010) 979--986.

\bibitem{hallock2004forecasting}
J.~L. Hallock~Jr, P.~J. Tharakan, C.~A. Hall, M.~Jefferson, W.~Wu, Forecasting
  the limits to the availability and diversity of global conventional oil
  supply, Energy 29~(11) (2004) 1673--1696.

\bibitem{barazandeh2016effect}
B.~Barazandeh, M.~Rafieisakhaei, A.~Moosavi, K.~Bastani, Effect of localization
  on the car market under intense sanctions; a system dynamics approach, in:
  The 34rd International Conference of the System Dynamics Society, 2016.

\bibitem{13_financialtimes_2015}
G.~Chazan,
  \href{http://www.ft.com/cms/s/0/0d6ac7c0-cbfb-11e1-aac1-00144feabdc0.html}{Oil
  demand shifts to emerging markets - ft}, Financial Times.
\newline\urlprefix\url{http://www.ft.com/cms/s/0/0d6ac7c0-cbfb-11e1-aac1-00144feabdc0.html}

\bibitem{basher2006oil}
S.~A. Basher, P.~Sadorsky, Oil price risk and emerging stock markets, Global
  finance journal 17~(2) (2006) 224--251.

\bibitem{morck2000information}
R.~Morck, B.~Yeung, W.~Yu, The information content of stock markets: why do
  emerging markets have synchronous stock price movements?, Journal of
  financial economics 58~(1-2) (2000) 215--260.

\bibitem{rafieisakhaei2016modeling}
M.~Rafieisakhaei, B.~Barazandeh, Modeling dynamics of a market-based emission
  control system: Efficacy analysis, in: Technologies for Sustainability
  (SusTech), 2016 IEEE Conference on, IEEE, 2016, pp. 213--218.

\bibitem{jaffe2002beijing}
A.~M. Jaffe, S.~W. Lewis, Beijing's oil diplomacy, Survival 44~(1) (2002)
  115--134.

\bibitem{baumeister2015understanding}
C.~Baumeister, L.~Kilian, Understanding the decline in the price of oil since
  june 2014, CFS Working Paper.

\bibitem{encykey}
IEA, Key world energy statistics, International Energy Agency.

\bibitem{economides2009state}
M.~J. Economides, D.~A. Wood, The state of natural gas, Journal of Natural Gas
  Science and Engineering 1~(1) (2009) 1--13.

\bibitem{rafieisakhaei2016supply}
M.~Rafieisakhaei, B.~Barazandeh, A.~Moosavi, M.~Fekri, K.~Bastani, Supply and
  demand dynamics of the oil market: A system dynamics approach, in: The 34rd
  International Conference of the System Dynamics Society, 2016.

\bibitem{kilian2009not}
L.~Kilian, Not all oil price shocks are alike: Disentangling demand and supply
  shocks in the crude oil market, American Economic Review 99~(3) (2009)
  1053--69.

\bibitem{krichene2002world}
N.~Krichene, World crude oil and natural gas: a demand and supply model, Energy
  economics 24~(6) (2002) 557--576.

\bibitem{Raf15}
M.~Rafieisakhaei, B.~Barazandeh, M.~Bolursaz, M.~Assadzadeh, Modeling dynamics
  of expectations on global oil price, in: The 33rd International Conference of
  the System Dynamics Society, MA, US, 2015.

\bibitem{azadeh2016unique}
A.~Azadeh, M.~Fekri, S.~Asadzadeh, B.~Barazandeh, B.~Barrios, A unique
  mathematical model for maintenance strategies to improve energy flows of the
  electrical power sector, Energy Exploration \& Exploitation 34~(1) (2016)
  19--41.

\bibitem{7_africaecon.org_2015}
Africaecon,
  \href{http://www.africaecon.org/index.php/africa_business_reports/read/73}{Africa
  economic institute : Nigeria and oil smuggling}, Africa Economic Institute.
\newline\urlprefix\url{http://www.africaecon.org/index.php/africa_business_reports/read/73}

\bibitem{gately1984ten}
D.~Gately, A ten-year retrospective: Opec and the world oil market, Journal of
  Economic Literature (1984) 1100--1114.

\bibitem{11_almonitor_2015}
F.~Taştekin,
  \href{http://www.al-monitor.com/pulse/originals/2014/04/turkey-syria-borders-smuggling-guns-conflict-kurds-pkk-isis.html#}{Turkey's
  syria borders an open door for smugglers, al-monitor: the pulse of the middle
  east}, Al Monitor.
\newline\urlprefix\url{http://www.al-monitor.com/pulse/originals/2014/04/turkey-syria-borders-smuggling-guns-conflict-kurds-pkk-isis.html#}

\bibitem{herbert2014partisans}
M.~Herbert, Partisans, profiteers, and criminals: Syria's illicit economy,
  Fletcher F. World Aff. 38 (2014) 69--213.

\bibitem{birol2010world}
F.~Birol, World energy outlook 2010, International Energy Agency.

\bibitem{sterman1988modeling}
J.~D. Sterman, G.~P. Richardson, P.~Davidsen, Modeling the estimation of
  petroleum resources in the united states, Technological Forecasting and
  Social Change 33~(3) (1988) 219--249.

\bibitem{18_union_of_concernedscientists_2015}
UCS,
  \href{http://www.ucsusa.org/clean_vehicles/smart-transportation-solutions/vehicle-policy/current-policies-and-legislation/benefits-of-reducing-us-oil-use.html#.VQEmBvxze69}{Fueling
  a better future (2013)}, Union of Concerned Scientists.
\newline\urlprefix\url{http://www.ucsusa.org/clean_vehicles/smart-transportation-solutions/vehicle-policy/current-policies-and-legislation/benefits-of-reducing-us-oil-use.html#.VQEmBvxze69}

\bibitem{19_mccain_2015}
B.~McCain,
  \href{http://www.forbes.com/sites/brucemccain/2015/02/09/the-facts-behind-oils-price-collapse/}{The
  facts behind oil's price collapse}, Forbes.
\newline\urlprefix\url{http://www.forbes.com/sites/brucemccain/2015/02/09/the-facts-behind-oils-price-collapse/}

\bibitem{1_barton_2015}
M.~Barton,
  \href{http://www.bloomberg.com/news/videos/2015-02-05/is-oil-price-volatility-the-new-norm-}{Is
  oil price volatility the new norm?}, Bloomberg.
\newline\urlprefix\url{http://www.bloomberg.com/news/videos/2015-02-05/is-oil-price-volatility-the-new-norm-}

\bibitem{2_webb_2015}
T.~Webb,
  \href{http://www.thetimes.co.uk/tto/business/industries/naturalresources/article4294934.ece}{Oil
  price volatility `is the new normal'}, The Times.
\newline\urlprefix\url{http://www.thetimes.co.uk/tto/business/industries/naturalresources/article4294934.ece}

\bibitem{20_conger_2015}
C.~Conger,
  \href{http://auto.howstuffworks.com/fuel-efficiency/vehicles/lithium-ion-battery-car.htm}{Will
  lithium-ion batteries power cars? - howstuffworks}, HowStuffWorks.
\newline\urlprefix\url{http://auto.howstuffworks.com/fuel-efficiency/vehicles/lithium-ion-battery-car.htm}

\bibitem{imf2015world}
IMF, World economic outlook, IMF.org.

\bibitem{15_bartash_2015}
J.~Bartash,
  \href{http://www.marketwatch.com/story/five-predictions-for-us-economy-in-2015-2014-12-31}{Five
  predictions for u.s. economy in 2015}, MarketWatch.
\newline\urlprefix\url{http://www.marketwatch.com/story/five-predictions-for-us-economy-in-2015-2014-12-31}

\bibitem{14_marketrealistcom_2015}
G.~Kristopher,
  \href{http://marketrealist.com/2015/01/economic-crisis-affects-price-crude-oil/}{How
  an economic crisis affects the price of crude oil - market realist},
  Marketrealist.com.
\newline\urlprefix\url{http://marketrealist.com/2015/01/economic-crisis-affects-price-crude-oil/}

\bibitem{22_the_economist_2015}
Economist, Let there be light,
  \url{http://www.economist.com/printedition/2015-01-17} (Jan 2015).

\bibitem{1_company_2013}
D.~J.~. Company, Spot oil price: West texas intermediate (discontinued series)
  (oilprice),
  \url{https://research.stlouisfed.org/fred2/series/OILPRICE/downloaddata},
  [Online; accessed 13-July-2015] (2015).

\bibitem{hamilton2009causes}
J.~D. Hamilton, Causes and consequences of the oil shock of 2007-08, Tech.
  rep., National Bureau of Economic Research (2009).

\bibitem{1_smith_2015}
J.~Smith,
  \href{http://www.rff.org/blog/2009/2008-oil-price-shock-markets-or-mayhem}{The
  2008 oil price shock: Markets or mayhem? | resources for the future} (2015).
\newline\urlprefix\url{http://www.rff.org/blog/2009/2008-oil-price-shock-markets-or-mayhem}

\end{thebibliography}

\end{document}